\documentclass[aps,prl,twocolumn,showpacs,preprintnumbers,amsmath,superscriptaddress,amssymb,floats,nofootinbib]{revtex4}

\usepackage{graphicx}
\usepackage{dcolumn}

\begin{document}

\def\Journal#1#2#3#4{{#1} {\bf #2}, #3 (#4)}

\def\NCA{\rm Nuovo Cimento}
\def\NIM{\rm Nucl. Instrum. Methods}
\def\NIMA{{\rm Nucl. Instrum. Methods} A}
\def\NPB{{\rm Nucl. Phys.} B}
\def\PLB{{\rm Phys. Lett.}  B}
\def\PRL{\rm Phys. Rev. Lett.}
\def\PRD{{\rm Phys. Rev.} D}
\def\PRC{{\rm Phys. Rev.} C}
\def\ZPC{{\rm Z. Phys.} C}	
\def\JPG{{\rm J. Phys.} G}
\def\st{\scriptstyle}
\def\sst{\scriptscriptstyle}
\def\mco{\multicolumn}
\def\epp{\epsilon^{\prime}}
\def\vep{\varepsilon}
\def\ra{\rightarrow}
\def\ppg{\pi^+\pi^-\gamma}
\def\vp{{\bf p}}
\def\ko{K^0}
\def\kb{\bar{K^0}}
\def\al{\alpha}
\def\ab{\bar{\alpha}}
\def\be{\begin{equation}}
\def\ee{\end{equation}}
\def\bea{\begin{eqnarray}}
\def\eea{\end{eqnarray}}
\def\CPbar{\hbox{{\rm CP}\hskip-1.80em{/}}}

\title{\large \bf A High-Pressure Polarized $^3$He Gas Target for Nuclear Physics Experiments Using A Polarized Photon Beam}

\author{Q. Ye, G. Laskaris, H. Gao, W. Chen, W. Zheng, X. Zong \\
{\it Triangle Universities Nuclear Laboratory and \\
Department of Physics, Duke University, Durham, NC 27708,~USA}\\
T. Averett\\
{\it 
Department of Physics, 
College of William and Mary, Williamsburg, VA ,~USA}\\
G. D. Cates, W. A. Tobias\\
{\it Department of Physics, University of Virginia, Charlottesville, VA 22904, ~USA}
}

\begin{abstract}
Following the first experiment on three-body photodisintegration of polarized $^3$He utilizing circularly polarized photons from High Intensity Gamma Source (HI$\gamma$S) at Duke Free Electron Laser Laboratory (DFELL), a new high-pressure polarized $^3$He target cell made of pyrex glass coated with a thin layer of sol-gel doped with aluminum nitrate nonahydrate has been built in order to reduce the photon beam induced background. The target is based on the technique of spin-exchange optical pumping of hybrid rubidium and potassium and the highest polarization achieved is $\sim$62\% determined from both NMR-AFP and EPR polarimetry. The $X$ parameter is estimated to be $\sim0.06$ and the performance of the target is in good agreement with theoretical predictions. We also present beam test results from this new target cell and the comparison with the GE180 $^3$He target cell used previously at HI$\gamma$S. This is the first time that sol-gel coating technique has been used in a polarized $^3$He target for nuclear physics experiments. 
\end{abstract}

\maketitle

\section{I. Introduction}

Quantum chromodynamics (QCD) is the theory of strong interaction in terms of quark and gluon degrees of freedom. While QCD has been well tested in the high energy regime where perturbative calculations can be carried out, it is still unsolved in the low energy, non-perturbative regime. 
Understanding the structure of the nucleon from the underlying theory of QCD, a fundamental and challenging task in nuclear and particle physics, 
remains an area of active research. With developments in polarized beam, 
recoil polarimetry, and polarized target technologies, 
polarization experiments have provided new observables  
on quantities related to the nucleon structure.

The High Intensity Gamma Source (HI$\gamma$S) at the Duke Free Electron Laser Laboratory (DFELL) opens a new window to studies of fundamental quantities related to the structure of the nucleon through polarized Compton scattering from polarized targets~\cite{higsreview}. Such measurements allow access to nucleon spin polarizabilities, which describe the response of a spin-aligned nucleon to a quasi-static external electromagnetic field. 
Since there are no stable free neutron targets, effective neutron targets, such as deuteron and $^3$He, are commonly used.
A polarized $^3$He target is an effective polarized neutron target~\cite{Friar,Blankleider} due to the fact that the neutron is
$\sim$90\% polarized in a polarized $^3$He nucleus. There have been extensive studies employing polarized $^3$He targets to extract the neutron electromagnetic form factors~\cite{Gao,Xu,Meyer,Becker,Rohe,Bermuth}, and neutron spin structure functions~\cite{Anthony,Abe,Ackerstaff, Zheng,e94010}. To extract information 
on neutron using a polarized $^3$He target, nuclear corrections need to be applied which rely on the state-of-the-art calculations of three-body systems.

The HI$\gamma$S facility also provides unique opportunities to test the three-body calculations. In 2008, a first measurement of double polarized three-body photodisintegration of $^3$He was carried out at HI$\gamma$S~\cite{zong} with an 
incident gamma beam energy of 11.4 MeV. In addition to providing tests of three-body calculations, three-body photodisintegration of $^3$He is of further 
importance to experimental tests of the Gerasimov-Drell-Hearn (GDH) sum rule~\cite{Drell}.
In this experiment, a high-pressure, longitudinally polarized $^3$He gas target~\cite{Kramer} and a circularly polarized photon beam were employed. Seven liquid scintillator detectors were placed around the $^3$He target to detect the neutrons 
from the three-body breakup channel. The $^3$He gas target cell was made of aluminosilicate (GE180) glass. This type of glass has fewer magnetic impurities and is less permeable to $^3$He atoms than regular pyrex glass. However, the rich concentration of barium in the GE180 glass produced a large amount of background events in the neutron 
detectors. To reduce the background for future measurements at HI$\gamma$S, a new high-pressure $^3$He cell made of sol-gel coated pyrex glass has been developed and tested. 

The coating technique was developed by doping sol-gel with aluminum nitrate nonahydrate (Al(NO$_3$)$_3\cdot$9H$_2$O)~\cite{hsu,Tobias}. This method produces a glass with better homogeneity
and higher purity via a chemical route. Single sealed pyrex cells produced using the sol-gel coating technique have yielded longer relaxation times than those from cells without the coating~\cite{hsu}. This is the first time that this technique has been applied to a high-pressure $^3$He target, a double-cell system for nuclear physics experiments. The smooth paramagnetic-free aluminosilicate glass coated surface reduces the probability of $^3$He depolarization from the wall. Its low $^3$He permeability helps prevent the loss of $^3$He atoms. This allows long-term operation at temperatures typical of spin exchange optical pumping process (185$^{\circ}$C for Rb-only-cells and 238$^{\circ}$C for Rb-K hybrid cells). The target cell ``BOLT'' was coated at the University of Virginia and filled at the College of William and Mary. A photon beam test of BOLT at HI$\gamma$S was carried out in May, 2009.
The rest of the paper is organized as the following.
Section II describes the experimental apparatus and procedure.
The target performance and a comparison between theoretical calculations and experimental results are presented in Section III. The in-beam test results of this new target are reported in Section IV.

\section{II. Experimental Apparatus and Procedure}

\begin{figure}[ht!]
\begin{center}
\includegraphics*[width=8cm]{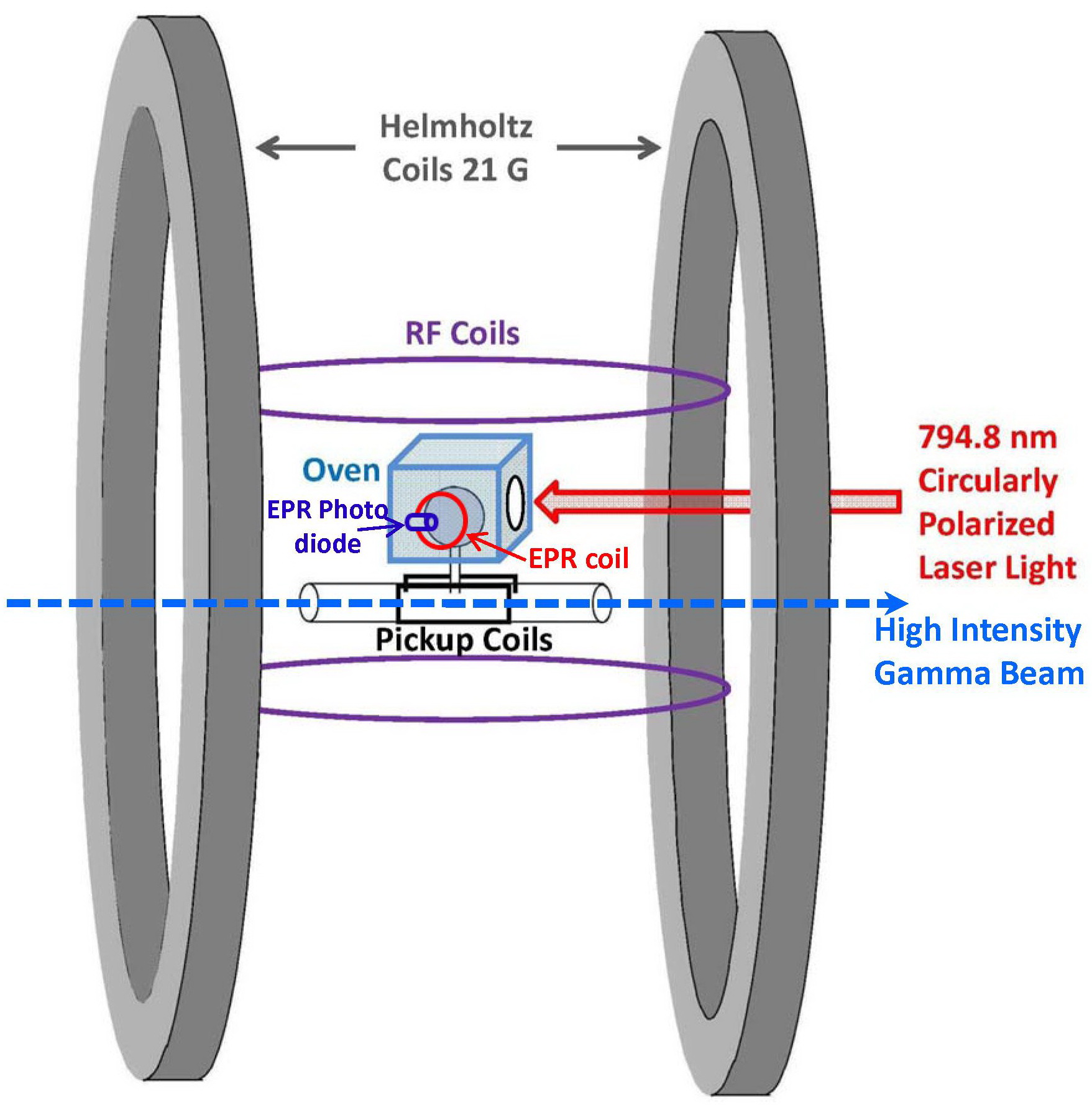}
\caption{(Color online) A schematic of the experimental setup.}
\label{experimentalschematics}
\end{center}
\end{figure}

A schematic of the experimental apparatus is shown in Fig.~\ref{experimentalschematics}.
It consists of a pair of Helmholtz coils with a diameter of 173 cm to 
provide a magnetic holding field with a typical value of 21 G.
BOLT is a pyrex glass cell coated with aluminosilicate and contains a mixture of Rb-K. It consists of a spherical pumping chamber with a radius of 4.3 cm and a target chamber with a length and a diameter of 38.7 cm and 3.1 cm, respectively. The chambers are connected by a tube
that is 9 cm long with a diameter of 1.3 cm. The cell is installed in the center of the Helmholtz coils with the pumping chamber in an oven.

The $^3$He polarization is measured using both the NMR-adiabatic fast passage (AFP) method~\cite{Lorenzon} and the electron paramagnetic resonance (EPR) technique~\cite{RomalisEPR}.
The AFP system includes two $\sim$79 cm-diameter RF coils with a separation of 39.5 cm placed horizontally above and below the cell and a pair of rectangular pickup coils located on both sides of the target chamber. The pickup coils are perpendicular to both the holding field and the RF field. The EPR system includes an 5.1 cm diameter EPR coil inside the oven close to the pumping chamber and a photo diode to monitor the EPR signal. Details of the polarimetry systems can be found in~\cite{Kramer}. The $^3$He nuclei are polarized through spin-exchange optical pumping.
Limits of alkali polarization have been observed for broad band laser light~\cite{Chann} and a spectrally narrowed laser is added to the experimental setup.
Three lasers with three separate sets of optics are used to produce circularly polarized laser light. After the optics, the net output power of the two Coherent DUO-FAP broad-band lasers is $\sim$78 W and the third Spectra-Physics narrowed laser has a net power output of 23 W.

Before the $^3$He nuclei are polarized, the cell is heated to 120$^{\circ}$C in a separate oven and a tunable laser is used to probe the line shape of the Rb D$_1$ transition. Collisions between Rb and
$^3$He can broaden the resonance lines of rubidium so that the width is proportional to the density of $^3$He in the cell~\cite{Romalis}. By measuring the width of the D$_1$ line, the density of $^3$He is determined to be 5.16$\pm$0.29 amagats.
To polarize the $^3$He nuclei, the pumping chamber is heated up in the oven by air flowing through three heaters. The pressure inside the cell is $\sim$7.66 atms with the pumping chamber at $(238\pm0.5)^{\circ}$C and target chamber at $(60\pm0.5)^{\circ}$C. 
The Rb atoms in the pumping chamber are polarized through the optical pumping process and then transfer angular momentum to the K atoms. The spin exchange collisions between K and $^3$He and between Rb and $^3$He subsequently polarize the $^3$He nuclei~\cite{hybridSEOP}. The time to reach the maximum polarization for such a Rb-K hybrid cell is much faster than a Rb-only cell due to the higher efficiency for polarizing $^3$He by K~\cite{Baranga}.

The $^3$He polarization measured by the NMR-AFP method is recorded every three hours during the polarization accumulation period (``pump-up'' period). After the polarization has reached a maximum, EPR measurements are carried out to measure the absolute $^3$He polarization, which can be compared to the value from the NMR signal after water calibration~\cite{Lorenzon}. 
With the lasers off and the alkali no longer in vapor form, 
AFP measurements are continued in order to measure the spin-lattice relaxation time, $T_1$, in the cell at room temperature.

The systematic error of the relaxation time is dominated by the uncertainty in the determination of the AFP losses, which is derived by fitting $n$ consecutive AFP measurements to $A_0 (1-L)^n$, where $L$ is the AFP inefficiency parameter. The relaxation times are obtained from the exponential decay of the AFP signal strength as a function of time, corrected for the AFP losses. The AFP system uses an RF amplifier to power the RF coils for the spin flip during the measurement, but the RF amplifier is left off during the time intervals between AFP measurements to prevent the polarized $^3$He atoms from depolarizing due to the wide band amplified RF noise. The AFP signals during the pump-up period are fitted to $A_0 (1-e^{-t/T_p})$ with the AFP inefficiency taken into account, where $T_p$ is the effective pump-up time for the double-cell system.

\section{III. Results and Discussion}

The three major contributions to $^3$He depolarization are the $^3$He dipole-dipole relaxation mechanism, the magnetic field gradient effect and the surface effect on the cell wall. The dipole-dipole relaxation
time is calculated to be $\sim$142 hours for BOLT~\cite{Newbury}. The relaxation time due to the magnetic field gradient in our system is calculated to be over 500 hours~\cite{CatesSchaefer} from the magnetic field mapping data.
The measured pump-up times, $T_p$, maximum NMR and EPR signals, and the double-cell room temperature relaxation times, $T_1$, are listed in Table~\ref{BOLTresult} together with the statistical uncertainties. Since the measured values for $T_1$ are much shorter than the relaxation times from the dipole-dipole effect and the gradient effect, the wall effect on the cell's surface is the most significant contribution to the relaxation of polarized $^3$He.
\begin{table}[!ht]
\begin{center}
\caption{Pump-up time ($T_p$), maximum NMR signal, maximum EPR signal, relaxation time ($T_1$) from different measurements for BOLT cell, statistical uncertainties included.}
\label{BOLTresult}
\vspace*{10pt}
\begin{tabular}{|c|c|c|c|c|}
\hline
Run no.& $T_p$ (hr) & NMR(mV) & EPR(kHz) & $T_1$(hr)\\
\hline
1 & 9.34$\pm$0.27 & 17.31$\pm$0.20 & 39.12$\pm$0.32 & 34.40$\pm$0.31\\
\hline
2 & 8.38$\pm$0.17 & 17.71$\pm$0.22 & 38.57$\pm$0.27 & 33.59$\pm$0.12\\
\hline
3 & 8.80$\pm$0.21 & 17.47$\pm$0.21 & 40.95$\pm$0.27 & 33.00$\pm$0.04\\
\hline
\end{tabular}
\end{center}
\end{table}
After comparing with the water NMR signal~\cite{Lorenzon}, the absolute $^3$He polarization in the target chamber is determined to be (60.9$\pm$4.1)\%. 
Since the EPR constant~\cite{RomalisEPR} for our cell is 1.57$\pm$0.09, the measured EPR frequency shift corresponds to a $^3$He polarization of (62.1$\pm$3.6)\%, which is consistent with the result from the NMR-AFP measurements after the water calibration. The uncertainty is dominated by the uncertainty in the $^3$He number density of the cell. This polarization is significantly higher than the highest 
target polarization ($\sim$46\% from cell ``ELVIS'') reported previously by our group from a hybrid 
cell~\cite{Kramer}.

In order to calculate the theoretically achievable maximum $^3$He polarization in BOLT, the alkali metal's polarization is estimated first. The Rb's polarization at any given location $z$ along the laser beam is~\cite{ChenGentile}
\begin{equation}
P_{Rb}(z,P)=\frac{R(z,P)}{R(z,P)+\Gamma^{'}_{Rb}}
\label{eq1}
\end{equation}
where $R(z,P)=\int \phi(z,\nu,P)\sigma_{Rb}(\nu)d\nu$ is the optical pumping rate with a certain laser power $P$, and $\Gamma^{'}_{Rb}$ is the total spin destruction rate. $\phi(z,\nu,P)$ is the number of photons per unit area $A$ per unit time in the frequency interval $d\nu$ ($d\nu=\frac{c}{\lambda^2}\delta\lambda$) along the laser beam direction at a given $z$ and $\nu$.
In our case, $A$=48.3 cm$^{2}$ and the FWHM of the broad band diode-laser is $\delta\lambda\sim4$ nm. $\sigma_{Rb}(\nu)=\frac{\sigma_{Rb0}}{1+4\Delta^{2}/\gamma_{Rb}^{2}}$~\cite{ChenGentile} is the frequency interval absorption cross section, assuming a Lorentzian form, where the peak cross section is $\sigma_{Rb0}=3.2\times 10^{-13}$ cm$^{2}$. $\Delta=\nu-\nu_{0}$ is the frequency offset from the optical pumping resonant frequency $\nu_{0}$ ($\lambda_{0}$=794.7 nm), and $\gamma_{Rb}=(18.7\pm0.3)$ GHz/amg~\cite{Romalis} is the pressure broadened width of the D$_1$ transition of Rb. $\phi(z,\nu,P)$ can be found by solving the equation~\cite{RomalisEPR,ChenGentile}
 \begin{equation}
 \frac{\partial\phi(z,\nu,P)}{\partial z}=-\phi(z,\nu,P)\sigma_{Rb}(\nu)[Rb](1-P_{\infty}P_{Rb}(z,P))
 \label{eq22}
 \end{equation}
Eqn. (1) and (2) present a highly nonlinear behavior and
a moderate approximation of Rb's polarization is made. By substituting the $P_{Rb}(z,P)$ with the $P_{Rb}(z_{max},P)$ at the back of the cell ($z\sim8$ cm) from the numerical simulation in~\cite{jadeep} with a $D=[K]/[Rb]\sim5$, an analytical expression of $\phi(z,\nu,P)$ can be derived (see Appendix I)
\begin{equation}
\phi(z,\nu,P)\cong\frac{P}{A h\nu_{0} d\nu}e^{-[Rb]\sigma_{Rb}(\nu)\big(1-P_{\infty}P_{Rb}(z_{max},P)\big)z}
\label{eq2}
\end{equation}
where $[Rb]$ is the Rb number density and $P_{\infty}$ is the mean photon spin of the pumping light. It follows that the pumping rate at any given location along $z$ is given by (see Appendix I)
\begin{eqnarray}
R(z,P)&\cong&\varepsilon\frac{\pi \sigma_{Rb0}\gamma_{Rb}P}{2 A h\nu_{0} d\nu}e^{-[Rb]\sigma_{Rb0}\big(1-P_{\infty}P_{Rb}(z_{max},P)\big)z/2}\nonumber \\
&&\cdot I_{0}\Big([Rb]\sigma_{Rb0}\big(1-P_{\infty}P_{Rb}(z_{max},P)\big)z/2\Big)
\label{eq3}
\end{eqnarray}
where $I_{0}$ is the zeroth order Bessel function of the first kind and $\varepsilon\sim0.75$ is an experimental parameter associated with the laser power loss due to the oven window and the cell's glass wall. Eqn. (4) gives a conservative estimation of the pumping rate and consequently the $P_{Rb}$.
It also suggests that Rb atoms are polarized to different degrees along the laser direction in the pumping chamber. By substituting eqn. (4) into eqn. (1) and taking the volume average, the $<P_{Rb}>$ can be found numerically for each laser power value $P$. The total spin destruction rate $\Gamma^{'}_{Rb}=\Gamma_{Rb}+D\Gamma_{K}+q_{KR}[K]$~\cite{ChenGentile} is calculated to be $\sim$1.3 kHz with
the nitrogen pressure $P_{N_{2}}=75$ torr at room temperature. $q_{KR}=2.2\times10^{-13}$ cm$^3$/s is the geometric mean of the Rb-Rb and K-K spin destruction rates. With the measured laser power of 78 W after all the optics from the two broadband Coherent diode lasers in addition to the 23 W narrowed band laser, which has a FWHM of $\delta\lambda=0.4$ nm corresponding to 230 W of broad band laser in the ideal case, the average Rb polarization is calculated to be $<P_{Rb}>\sim88.0\%$, based on a conservative estimation.

\begin{figure}[ht!]
\begin{center}
\includegraphics*[width=8cm]{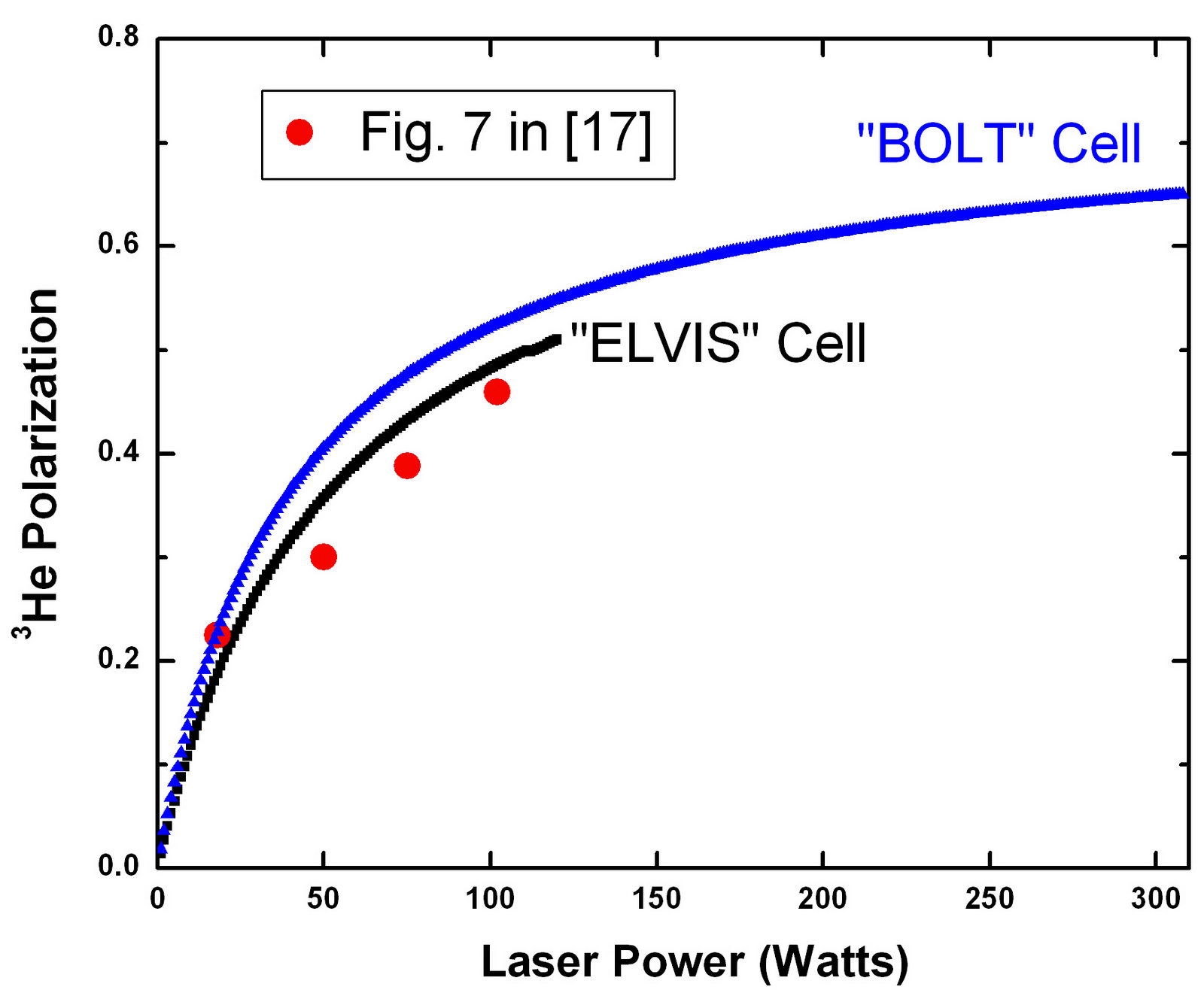}
\caption{(Color online) The theoretical $^3$He polarization versus the input broad band laser power for BOLT (this paper) and ELVIS (in~\cite{Kramer}). The red circles represent the four data points from Fig. 7 in~\cite{Kramer} and they are in good agreement with theoretical calculations for ELVIS.}
\label{3HePvsLaserE}
\end{center}
\end{figure}

After the Rb polarization is calculated, the polarization of $^3$He can be estimated by employing eqn. (A43) in~\cite{ChuppLoveman}. This equation needs to be modified according to eqn. (2) in~\cite{Chann} (or eqn. (1) in~\cite{Babcock}) in order to include the $X$ parameter which is a phenomenological parameter that reflects additional unknown spin relaxation processes. 
Assuming the $^3$He transfer rate between the pumping chamber and the target chamber is much higher than the spin relaxation rate, the $^3$He polarization in both the pumping and target chambers is given by (see Appendix II)
\begin{equation}
P_{3He}=<P_{Rb}(P)>\frac{f_{opc}\cdot\gamma^{Rb/K}_{se}}{f_{opc}\cdot\gamma^{Rb/K}_{se}(1+X)+1/T_{1}}
\label{eq4}
\end{equation}
where $\gamma^{Rb/K}_{se}$ is the spin exchange rate of Rb and K with $^3$He and $f_{opc}=0.38$ is the fraction of $^3$He atoms in the pumping chamber. Since $T_p=\frac{1}{\Gamma_{3He}}$ and $\Gamma_{3He}=f_{opc}\cdot\gamma^{Rb/K}_{se}(1+X)+1/T_{1}$~\cite{Babcock}, we have
\begin{eqnarray}
\gamma^{Rb/K}_{se}&=&\frac{P_{3He}}{f_{opc}\cdot <P_{Rb}>\cdot T_{p}} \label{eq5}\\
\frac{1}{T_{p}}&=&f_{opc}\cdot\gamma_{se}^{Rb/K}(1+X)+1/T_{1}
\label{eq6}
\end{eqnarray}
Using the experimental value of pump-up time, the $^3$He polarization and the relaxation time, the $X$ parameter can be estimated using the above equations to be $\sim$0.06.
The theoretical spin exchange rate is given by $\gamma^{Rb/K}_{se}=k^{Rb}_{se}[Rb]+k^{K}_{se}[K]=7.4\times10^{-5}$ Hz~\cite{ChenGentile,Babcock}, where $k^{Rb}_{se}=6.8\times10^{-20}$ cm$^{3}$/s~\cite{ChannBabcock,Baranga} and $k^{K}_{se}=5.5\times10^{-20}$ cm$^{3}$/s~\cite{BabcockThesis}.
The $^3$He polarization can then be calculated and plotted versus the input broad band laser power by inserting eqn. (\ref{eq1}) into eqn. (\ref{eq4}) along with the $X$ parameter and the theoretical spin exchange rate (Fig.~\ref{3HePvsLaserE}).
The $^3$He polarization in BOLT can reach a value of $P_{3He}=(65.0\pm2.9)$\% with an input broad band laser power of 78 W and 23 W narrowed band laser, which satisfactorily agrees with the experimental results.
If the exchange time is not neglected, eqn. (23) in Appendix II can be utilized to give a more precise estimation. From eqn. (2) in~\cite{Ye2009}, the relaxation rate $\Gamma=\frac{1}{T_1}=\frac{1}{4}\bar{v}\cdot DP\cdot \frac{S}{V}$, where $DP$ is the depolarization probability of a polarized $^3$He atom after each collision with the cell surface. Assuming the walls of the pumping chamber and the target chamber have the same DP, the relaxation rate is proportional to the product of mean $^3$He particle speed, surface to volume ratio and the total amount of $^3$He atoms in each chamber. $\Gamma_p$ and $\Gamma_t$ can be calculated using these known parameters and eqn. (23) then gives a $^3$He polarization of $\sim62.9$\% assuming $G_p=G_t=\frac{1}{100~\text{mins}}$, which agrees with the experimental results even better, though two assumptions are used in the above estimation.

The $^3$He polarization versus the broad band laser power is also calculated for the ELVIS cell from~\cite{Kramer} and plotted in the same figure. Only broad band lasers were used in~\cite{Kramer} and the total laser power went up to 120 W in the calculation. The experimental data (red circles) from Fig. 7 in~\cite{Kramer} are in good agreement with the theoretical calculations.
We believe the better performance of BOLT over ELVIS is mostly due to the following two factors: a spectrally narrowed laser used together with the broadband lasers, and a significantly lower $^3$He number density of the BOLT cell in comparison with that of the ELVIS cell. 

\section{IV. Beam test}

An in-beam test of BOLT was carried out at HI$\gamma$S in May, 2009. Three liquid organic scintillator detectors~\cite{Crowell} filled with BC-501A fluid from the Bicron Corporation were placed at 50$^\circ$, 90$^\circ$, 130$^\circ$ and 90 cm, 75 cm, 90 cm away from the center of the target, respectively.
A vertical multi-layer motor-controlled support was used to place different targets in the gamma beam. The targets included N$_2$ gas targets in both pyrex and GE180 glass cells, BOLT, and a GE180 glass cell ``Linda''. A thin aluminum plate, liquid D$_2$O target and D$_2$ gas targets were available for detector calibration purposes. All N$_2$ and $^3$He target cells have the same dimensions.

The energy of the photon beam was 11.4 MeV and the photon flux was $\sim 1.5\times 10^{7}$/s monitored by a three-paddle system~\cite{Blackston} right after the photon beam's collimator and a 4.7 cm thick liquid D$_2$O target placed downstream of the $^3$He target with two neutron detectors at 90$^\circ$ on both sides. Compared to the position in the experiment~\cite{zong}, the $^3$He target was pushed $\sim$2 meters downstream due to another experimental apparatus in the beamline.

\begin{figure}[ht!]
\begin{center}
\includegraphics*[width=8cm]{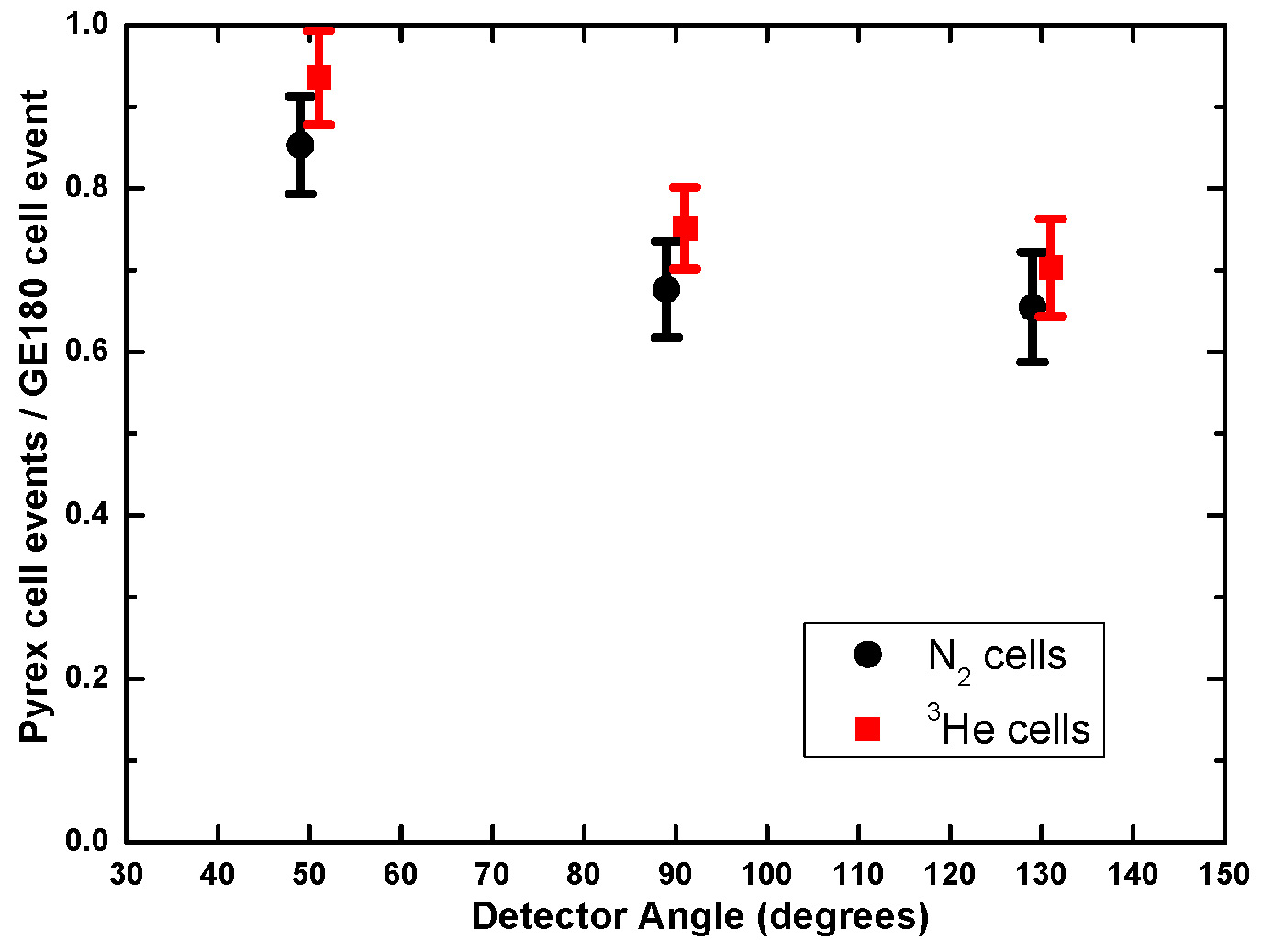}
\caption{(Color online) Results of the in-beam test. Black circles represent the ratios between the events from pyrex glass N$_2$ cell and GE180 glass N$_2$ cell at three detectors together with the statistical uncertainties. The red squares represent the ratios between the events from BOLT and Linda at three detectors together with the statistical uncertainties.}
\label{beamtest}
\end{center}
\end{figure}

Fig.~\ref{beamtest} shows the results of the in-beam test.
The points represent the neutron events' ratios between two N$_2$ cells made of pyrex glass and GE180 glass and two $^3$He cells made of pyrex (BOLT) and GE180 (Linda) glass. All the neutron events were selected from the raw data after the multiplicity cut, pulse shape discrimination cut, pulse height cut and time of flight cut~\cite{Blackston} and had energies ranging from 1 MeV to 2.5 MeV. The numbers are also shown in Table~\ref{beamtestResult}.

\begin{table}[!ht]
\begin{center}
\caption{Results of the in-beam test. The ratios between the number of events from N$_2$ and $^3$He cells made from pyrex and GE180 glass are listed.}
\label{beamtestResult}
\vspace*{10pt}
\begin{tabular}{|c|c|c|}
\hline
Detector & Ratio between N$_2$ cells & Ratio between $^3$He cells \\
\hline
50$^{\circ}$ & 85.3$\pm$6.0\% & 93.5$\pm$5.7\% \\
\hline
90$^{\circ}$ & 67.6$\pm$5.9\% & 75.2$\pm$5.0\% \\
\hline
130$^{\circ}$& 65.5$\pm$6.7\% & 70.3$\pm$6.0\% \\
\hline
\end{tabular}
\end{center}
\end{table}

Since the $^3$He target was placed farther downstream in this experiment, the liquid D$_2$O target and the longer air path generated considerably more background neutrons. In the 50$^{\circ}$ detectors, the yield difference is smaller since they are closer to the liquid D$_2$O target and the background contamination dilutes the difference.
The improvement using BOLT for future experiments compared with Linda should be better than the results presented in Fig.~\ref{beamtest} for an optimal target position, where the background neutron events can be minimized. Furthermore,
plans have been made to employ more stable, higher flux (at least $5\times10^{7}$/s) photon beams, better shielding from the downstream D$_2$O target and a vacuum pipe between the beam source and the target.
All these measures will help reduce the neutron background further for better statistics.

\section{V. Summary}

We have tested the world's first high-pressure hybrid $^3$He pyrex glass cell coated with a thin aluminosilicate glass and have achieved a polarization of $\sim$62\% determined by both NMR-AFP and EPR methods. This value compares favorably with the predicted maximum theoretical $^3$He polarization value.
The $X$ parameter is measured for the first time in a hybrid double-cell system to be $\sim0.06$.
The in-beam test shows that the sol-gel coated pyrex glass target generates fewer neutron background events than the GE180 glass cell. This new type of target will be important for the future Compton scattering and GDH experiments at HI$\gamma$S.

\section{VI. ACKNOWLEDGMENT}

The authors wish to thank Michael Souza of Princeton University
for making the target cell reported in this work, Alexandre Deur for providing us with chemical information for the GE180 glass,
K. Kluttz for helpful comments about this work, Yi Qiang and Yi Zhang for their help with the target density measurement, N. Brown, M. Emamian, S. Henshaw, B. Perdue and H. Weller for helping with the target alignment, M. Pentico, V. Rathbone, C. Sun and Y. Wu for the gamma beam operation, B. Carlin and C. Westerfeld for the technical support. This work is supported by the U.S. Department of Energy under contract number DE-FG02-03ER41231 and Duke University.

\section{Appendix I}

The calculation of $\phi(z,\nu,P)$ (eqn. (\ref{eq2})) assumes that the spectrum of the laser has a Gaussian form and the power $P$ is concentrated in the FWHM $\delta\lambda$. This is an approximation because a small part of the laser power lies in the tails of the Gaussian.

In order to find the optical pumping rate $R(z,P)$ (eqn. (\ref{eq3})), a second approximation is needed. Since
\begin{eqnarray}
R(z,P)&=&\int_{0} ^{\infty} \phi(z,\nu,P)\sigma_{Rb}(\nu)d\nu \nonumber \\
&=& \frac{P}{A h\nu_{0} d\nu} \int_{0}^{\infty} e^{-\xi\sigma_{Rb}(\nu)}\sigma_{Rb}(\nu)d\nu
\label{eq9}
\end{eqnarray}
where $\xi=[Rb](1-P_{\infty}P_{Rb}(z,P))z$. If the substitution $x=\frac{2\Delta}{\gamma_{Rb}}$ is made, the equation becomes
\begin{equation}
R(z,P)=\frac{\sigma_{Rb0}\gamma_{Rb}P}{2 A h\nu_{0} d\nu}\int_{-\frac{2\nu_{0}}{\gamma_{Rb}}} ^{\infty}e^{-\frac{\xi}{1+x^{2}}}\frac{1}{1+x^{2}}dx
\label{eq10}
\end{equation}
This integral is approximately equal to
\begin{equation}
R(z,P)\cong \frac{\sigma_{Rb0}\gamma_{Rb}P}{2 A h\nu_{0} d\nu} \int_{-\infty} ^{\infty}e^{-\frac{\xi}{1+x^{2}}}\frac{1}{1+x^{2}}dx
\label{eq11}
\end{equation}
for any numerical value of $\xi$. Since the integration function is symmetric w.r.t. 0, eqn. (\ref{eq11}) equals
\begin{equation}
R(z,P)\cong \frac{\sigma_{Rb0}\gamma_{Rb}P}{A h\nu_{0} d\nu} \int_{0} ^{\infty}e^{-\frac{\xi}{1+x^{2}}}\frac{1}{1+x^{2}}dx
\label{eq12}
\end{equation}
After a change of variable $y=\frac{1}{1+x^{2}}$ is made
\begin{equation}
R(z,P)\cong \frac{\sigma_{Rb0}\gamma_{Rb}P}{2 A h\nu_{0} d\nu} \int_{0} ^{1} \frac{1}{\sqrt{(1-y)y}}e^{-\xi y}dy
\label{eq13}
\end{equation}
Solving it gives eqn. (\ref{eq3}).

\section{Appendix II}

The rate equations for a double-cell system are given by (rate equations in~\cite{ChuppLoveman} with the $X$ parameter included)
\begin{widetext}
\begin{eqnarray}
\frac{dP_{p}}{dt}&=&-G_{p}(P_{p}-P_{t})-((1+X)f_{opc}\gamma_{se}^{Rb/K}+\Gamma_{p})P_{p}+<P_{Rb}>f_{opc}\gamma_{se}^{Rb/K} \\
\frac{dP_{t}}{dt}&=&G_{t}(P_{p}-P_{t})-\Gamma_{t}P_{t} \\
P_{p}(0)&=&P_{t}(0)=0; ~P'_{p}(\infty)=P'_{t}(\infty)=0
\end{eqnarray}
\end{widetext}
The assumptions are that the alkali metals are confined in the pumping chamber only and the sources of $^{3}$He polarization in the pumping chamber are spin exchange between alkali metals and $^3$He and diffusion of polarized $^3$He atoms from the target cell. Relaxation in the pumping (target) chamber is due to the combined relaxation mechanisms (dipole-dipole effect, magnetic field gradient effect and wall effect) and $^3$He diffusion to the target (pumping) chamber. The $G_{p}$ is the polarized $^3$He transfer rate from the pumping cell to the target cell and $G_{t}$ is the transfer rate in the other direction (Appendix in~\cite{ChuppLoveman}). They are defined as
\begin{eqnarray}
G_{p}&=&\frac{D_{p}S}{LV_{p}}\\
G_{t}&=&\frac{D_{t}S}{LV_{t}}
\end{eqnarray}
where $S$, $L$ are the cross-sectional area and length of the connecting tube between the pumping chamber and target chamber.
$D_{p(t)}$ and $V_{p(t)}$ are the $^3$He diffusion coefficient and volume of each chamber, respectively. The diffusion coefficient is given by $\frac{\overline{v}\lambda}{3}$ where $\overline{v}$ is the $^3$He mean thermal velocity and $\lambda$ is the mean free path. The total theoretical transfer rate is equal to $G=G_{p}+G_{t}$.

The relaxation rates are defined as (eqn. (6) in~\cite{ChuppLoveman})
\begin{eqnarray}
\Gamma_{p}&=&\Gamma'_{p}\frac{n_{p}V_{p}}{n_{p}V_{p}+n_{t}V_{t}}\\
\Gamma_{t}&=&\Gamma'_{t}\frac{n_{t}V_{t}}{n_{p}V_{p}+n_{t}V_{t}}\end{eqnarray}
where $\Gamma'_{p}$ and $\Gamma'_{t}$ are the averaged relaxation rates in the pumping and target chambers.
Theoretically, the spin exchange rate is given by $\gamma^{Rb/K}_{se}=k^{Rb}_{se}[Rb]+k^{K}_{se}[K]$~\cite{Babcock}.

The general time-dependent solution of this coupled equations system is given by
\begin{widetext}
\begin{eqnarray}
P_{p}(t) &=& \frac{(G_{t}+\Gamma_{t})<P_{Rb}>f_{opc}\gamma_{se}^{Rb/K}}{(G_{p}+f_{opc}\gamma_{se}^{Rb/K}(1+X)+\Gamma_{p})(G_{t}+\Gamma_{t})-G_{p}G_{t}}+ \frac{<P_{Rb}>f_{opc}\gamma_{se}^{Rb/K}}{\sqrt{(G_{p}+f_{opc}\gamma_{se}^{Rb/K}(1+X)+\Gamma_{p}-G_{t}-\Gamma_{t})^{2}+4G_{p}G_{t}}} \nonumber \\
& \cdot &
\Big[-\frac{G_{p}+f_{opc}\gamma_{se}^{Rb/K}(1+X)+\Gamma_{p}-G_{t}-\Gamma_{t}+\sqrt{(G_{p}+f_{opc}\gamma_{se}^{Rb/K}(1+X)+\Gamma_{p}-G_{t}-\Gamma_{t})^{2}+4G_{p}G_{t}}}{G_{p}+f_{opc}\gamma_{se}^{Rb/K}(1+X)+\Gamma_{p}+G_{t}+\Gamma_{t}+\sqrt{(G_{p}+f_{opc}\gamma_{se}^{Rb/K}(1+X)+\Gamma_{p}-G_{t}-\Gamma_{t})^{2}+4G_{p}G_{t}}} \nonumber \\ 
& \cdot & e^{-\frac{1}{2}(G_{p}+f_{opc}\gamma_{se}^{Rb/K}(1+X)+\Gamma_{p}+G_{t}+\Gamma_{t}+\sqrt{(G_{p}+f_{opc}\gamma_{se}^{Rb/K}(1+X)+\Gamma_{p}-G_{t}-\Gamma_{t})^{2}+4G_{p}G_{t}})t}\nonumber \\ & + &
\frac{G_{p}+f_{opc}\gamma_{se}^{Rb/K}(1+X)+\Gamma_{p}-G_{t}-\Gamma_{t}-\sqrt{(G_{p}+f_{opc}\gamma_{se}^{Rb/K}(1+X)+\Gamma_{p}-G_{t}-\Gamma_{t})^{2}+4G_{p}G_{t}}}{G_{p}+f_{opc}\gamma_{se}^{Rb/K}(1+X)+\Gamma_{p}+G_{t}+\Gamma_{t}-\sqrt{(G_{p}+f_{opc}\gamma_{se}^{Rb/K}(1+X)+\Gamma_{p}-G_{t}-\Gamma_{t})^{2}+4G_{p}G_{t}}}  \nonumber \\ 
& \cdot &
e^{-\frac{1}{2}(G_{p}+f_{opc}\gamma_{se}^{Rb/K}(1+X)+\Gamma_{p}+G_{t}+\Gamma_{t}-\sqrt{(G_{p}+f_{opc}\gamma_{se}^{Rb/K}(1+X)+\Gamma_{p}-G_{t}-\Gamma_{t})^{2}+4G_{p}G_{t}})t}\Big] 
\end{eqnarray}
\end{widetext}

\begin{widetext}
\begin{eqnarray}
P_{t}(t) &=& \frac{G_{t}<P_{Rb}>f_{opc}\gamma_{se}^{Rb/K}}{(G_{p}+f_{opc}\gamma_{se}^{Rb/K}(1+X)+\Gamma_{p})(G_{t}+\Gamma_{t})-G_{p}G_{t}}+\frac{2G_{t}<P_{Rb}>f_{opc}\gamma_{se}^{Rb/K}}{\sqrt{(G_{p}+f_{opc}\gamma_{se}^{Rb/K}(1+X)+\Gamma_{p}-G_{t}-\Gamma_{t})^{2}+4G_{p}G_{t}}} \nonumber \\
& \cdot &
\Big[\frac{1}{G_{p}+f_{opc}\gamma_{se}^{Rb/K}(1+X)+\Gamma_{p}+G_{t}+\Gamma_{t}+\sqrt{(G_{p}+f_{opc}\gamma_{se}^{Rb/K}(1+X)+\Gamma_{p}-G_{t}-\Gamma_{t})^{2}+4G_{p}G_{t}}} \nonumber \\ 
& \cdot &
e^{-\frac{1}{2}(G_{p}+f_{opc}\gamma_{se}^{Rb/K}(1+X)+\Gamma_{p}+G_{t}+\Gamma_{t}+\sqrt{(G_{p}+f_{opc}\gamma_{se}^{Rb/K}(1+X)+\Gamma_{p}-G_{t}-\Gamma_{t})^{2}+4G_{p}G_{t}})t}\nonumber \\ & - &
\frac{1}{G_{p}+f_{opc}\gamma_{se}^{Rb/K}(1+X)+\Gamma_{p}+G_{t}+\Gamma_{t}-\sqrt{(G_{p}+f_{opc}\gamma_{se}^{Rb/K}(1+X)+\Gamma_{p}-G_{t}-\Gamma_{t})^{2}+4G_{p}G_{t}}}  \nonumber \\ 
& \cdot &
e^{-\frac{1}{2}(G_{p}+f_{opc}\gamma_{se}^{Rb/K}(1+X)+\Gamma_{p}+G_{t}+\Gamma_{t}-\sqrt{(G_{p}+f_{opc}\gamma_{se}^{Rb/K}(1+X)+\Gamma_{p}-G_{t}-\Gamma_{t})^{2}+4G_{p}G_{t}})t}\Big]
\end{eqnarray}
\end{widetext}

Using condition $P'_{p}(\infty)=P'_{t}(\infty)=0$, the equilibrium $^3$He polarization in each chamber is

\begin{widetext}
\begin{eqnarray}
P_{p}(t\rightarrow \infty) &=& \frac{(G_{t}+\Gamma_{t})<P_{Rb}>f_{opc}\gamma_{se}^{Rb/K}}{(G_{p}+f_{opc}\gamma_{se}^{Rb/K}(1+X)+\Gamma_{p})(G_{t}+\Gamma_{t})-G_{p}G_{t}}\\
P_{t}(t\rightarrow \infty) &=& \frac{G_{t} <P_{Rb}>f_{opc}\gamma_{se}^{Rb/K}}{(G_{p}+f_{opc}\gamma_{se}^{Rb/K}(1+X)+\Gamma_{p})(G_{t}+\Gamma_{t})-G_{p}G_{t}}
\end{eqnarray}
\end{widetext}
The transfer rate is also measured experimentally by destroying the polarization in the target chamber using a rectangular RF coil, and measuring the recovery of the NMR free induction decay signal as polarized $^3$He atoms diffuse into the target chamber from the pumping chamber. The rate is measured to be $\sim\frac{1}{50~\text{mins}}$, which is much higher than the spin relaxation rate $\sim\frac{1}{33~\text{hrs}}$. So it is assumed that $G_{t}+\Gamma_{t}\simeq G_{t}$. And we have
\begin{widetext}
\begin{equation}
P_{p}(t\rightarrow \infty) = P_{t}(t\rightarrow \infty)=\frac{G_{t}<P_{Rb}>f_{opc}\gamma_{se}^{Rb/K}}{(G_{p}+f_{opc}\gamma_{se}^{Rb/K}(1+X)+\Gamma_{p})(G_{t}+\Gamma_{t})-G_{p}G_{t}}
\end{equation}
\end{widetext}
Assuming that $G_{p}\simeq G_{t}$, the equation becomes
\begin{equation}
P_{^{3}He}=<P_{Rb}>\frac{f_{opc}\gamma^{Rb/K}_{se}}{f_{opc}\gamma^{Rb/K}_{se}(1+X)+1/T_{1}}
\end{equation}
where $\Gamma=\frac{1}{T_{1}}=\Gamma_{p}+\Gamma_{t}$ is the total relaxation rate.

\end{document}